\documentclass[doublecol,a4paper,showpacs]{epl2}
\usepackage{graphicx}
\usepackage{amsmath}
\usepackage{amssymb}
\usepackage{enumerate}
\usepackage{subfigure}
\usepackage{tabularx}
\usepackage[colorlinks=true, pdfstartview=FitV, linkcolor=blue, citecolor=red, urlcolor=black, breaklinks=true]{hyperref}
\newcommand{\be}{\begin{equation}}
\newcommand{\ee}{\end{equation}}
\newcommand{\ben}{\begin{eqnarray}}
\newcommand{\een}{\end{eqnarray}}
\newcommand{\bes}{\begin{subequations}}
\newcommand{\ees}{\end{subequations}}
\def\bal#1\eal{\begin{align}#1\end{align}}

\title{Basin entropy behavior in a cyclic model of the rock-paper-scissors type}
\author{M. Mugnaine\inst{1} \and F.M. Andrade\inst{1,2} \and J.D. Szezech, Jr.\inst{1,2} \and D. Bazeia \inst{3}}
\shortauthor{M. Mugnaine \etal}
\institute{                    
  \inst{1} Graduate Program in Sciences, State University of Ponta Grossa, Ponta Grossa, Paran\'a, Brazil\\
  \inst{2} Department of Mathematics and Statistics, State University of Ponta Grossa, Ponta Grossa, Paran\'a, Brazil\\
  \inst{3} Departamento de F\'\i sica, Universidade Federal da Para\'\i ba, 58051-970 Jo\~ao Pessoa, Para\'\i ba, Brazil}


\abstract{We deal with stochastic network simulations in a model with three distinct species that compete under cyclic rules which are similar to the rules of the popular rock-paper-scissors game. We investigate the Hamming distance density and then the basin entropy behavior, running the simulations for some typical values of the parameters mobility, predation and reproduction and for very long time evolutions. The results show that the basin entropy is another interesting tool of current interest to investigate chaotic features of the network simulations that are usually considered to describe aspects of biodiversity in the cyclic three-species model.}

\begin{document}

\maketitle

\section{Introduction}

The popular rock-paper-scissors or RPS game is based on three simple rules that establish the behavior of rock, paper and scissors in the game: rock breaks scissors, scissors cut paper and paper wraps rock. It is a competition game and has been used to study important aspects of biodiversity in nature because it evolves cyclically and the cyclical features it engenders is important to keep biodiversity. The dynamics and stability of the cyclic three-species system depend on the specific interactions among the constituent species and in the past decade, some interesting works used the RPS rules to describe specific aspects of biodiversity; see, e.g., \cite{A,B,C} and references therein. 

In the investigations developed in \cite{A,B}, the authors studied competitive interactions between {\sl Escherichia coli} populations. In \cite{A}, they empirically tested a non-transitive model community containing three populations of {\sl Escherichia coli}. An interesting result was that diversity may be lost in experimental community when dispersal and interaction occur over relatively large spatial scales, whereas the populations coexist when ecological processes are localized. In \cite{B}, a new environment with the antibiotic-mediated antagonism was studied, and it was shown that coexistence occurred from a clumped spatial distribution of producers, suggesting that each producer can block invasion of the other producer. An agent-based simulation of the competition was considered, using the colicin version of the RPS model, in which the strains that produce colicins (C) kill sensitive (S) strains, which outcompete resistant (R) strains, which outcompete C strains. The study demonstrated that competitions between these three strains may lead the complete system in dynamic equilibrium, promoting microbial diversity in the environment. In \cite{C}, the authors focused on the variation of mobility in systems modelled by the standard rules of mobility, competition and reproduction, with competition described as in the cyclic RPS game. The results showed that mobility may promote or jeopardize biodiversity, depending on the value it gets, compared to competition and reproduction. 

There are many other investigations on biodiversity described under the RPS rules and generalisations to four and more species; some of them are reviewed in Refs.~\cite{review1,review2}. In the more recent work \cite{SR1}, a novel procedure to identify the chaotic behavior engendered by the network simulations used to describe a cyclic three-species system was developed. The procedure follows the Hamming distance concept, and it was used to provide a way to unveil the chaotic behavior of the time evolution that follows the rules of the RPS cyclic competition model. The subject was further examined in \cite{Ham}, to see how the Hamming distance density changes under modifications on the number of species and the size of the lattice. The two investigations showed that the Hamming distance density engenders universal qualitative behavior, which does not depend neither on the size of the lattice nor on the number of species. In the current work we further consider the case of three distinct species, but we consider two different square lattices, with sizes $200\times 200$ and $500\times 500$, since we want to add another type of simulation, that may also serve to infer the chaotic behavior of the stochastic network simulations that are generically used to describe the time evolution; see, e.g., Refs.~\cite{review1,review2,SR1,Ham} and references therein. In the next section we explain how the stochastic network simulations are implemented to describe the time evolution of the system.

The main motivation of the work is to develop another procedure to deal with the chaotic behavior engendered by the simulations that appear in games of the RPS type, and to verify how it goes along with the results on the Hamming distance that appeared before in \cite{SR1,Ham}. The subject is motivated by the fact that in \cite{SR2}, the authors proposed an interesting route to study complex systems via the basin entropy concept. This relies on a different framework, which was recently used in \cite{PRE} to study the barriers and transport through the phase space in nontwist systems. The basin entropy concept was also used to quantify the unpredictability of the final state in cold atoms experiments \cite{PRA}. In the current work, we want to apply the basin entropy concept to study the time evolution of the RPS model with three distinct species that compete cyclically. We also want to compare the results of the basin entropy with the results of the Hamming distance density, to see how they behave under the time evolution which we implement standardly.

Another inspection to be implemented concerns the increasing of the mobility and the ending of biodiversity. This was investigated before in Refs.~\cite{C,review2,Ham,E1,E2,E4,E5,M1,M2,M3}, and the authors showed that when the mobility increases toward a critical value, the system loses diversity, ending up with a single species. Inspired by this fact, here we also study how the basin entropy behaves for a very high value of the mobility, above the critical mobility.
 
To comply with the above motivations, we organize the work as follows. In the next section we introduce and explain all the steps required to perform the simulations and in the third section we briefly review the study on the Hamming distance density that appeared before in \cite{SR1,Ham}. We go on and in the fourth section we examine the basin entropy concept, and adapt it to the systems to be investigated in the current work. We end the work with comments and conclusions, and with some perspectives for future works.  

\section{Procedure}
We consider a model of a system of three distinct species $a, b$, and $c$ and use the colors red, blue, and yellow, respectively, to identify the set of species. In the model, the three species evolve in time under the rules of mobility $(m)$, reproduction $(r)$ and competition or predation
$(p)$, which are normalized to obey $m+p+r=1$; also, for simplicity we consider $p=r$, since distinct values for $p$ and $r$ do not qualitatively change the issues to be investigated in this work.  We use square lattices of size $L=N\times N$, with $N=200$, $500$, and also $1000$, which obey periodic boundary conditions. Also, we consider the Moore neighborhood, where any site in the lattice has eight neighbors, two horizontally, two vertically and four diagonally.

\begin{figure}[t!]
\centering
\includegraphics[width=0.98\linewidth]{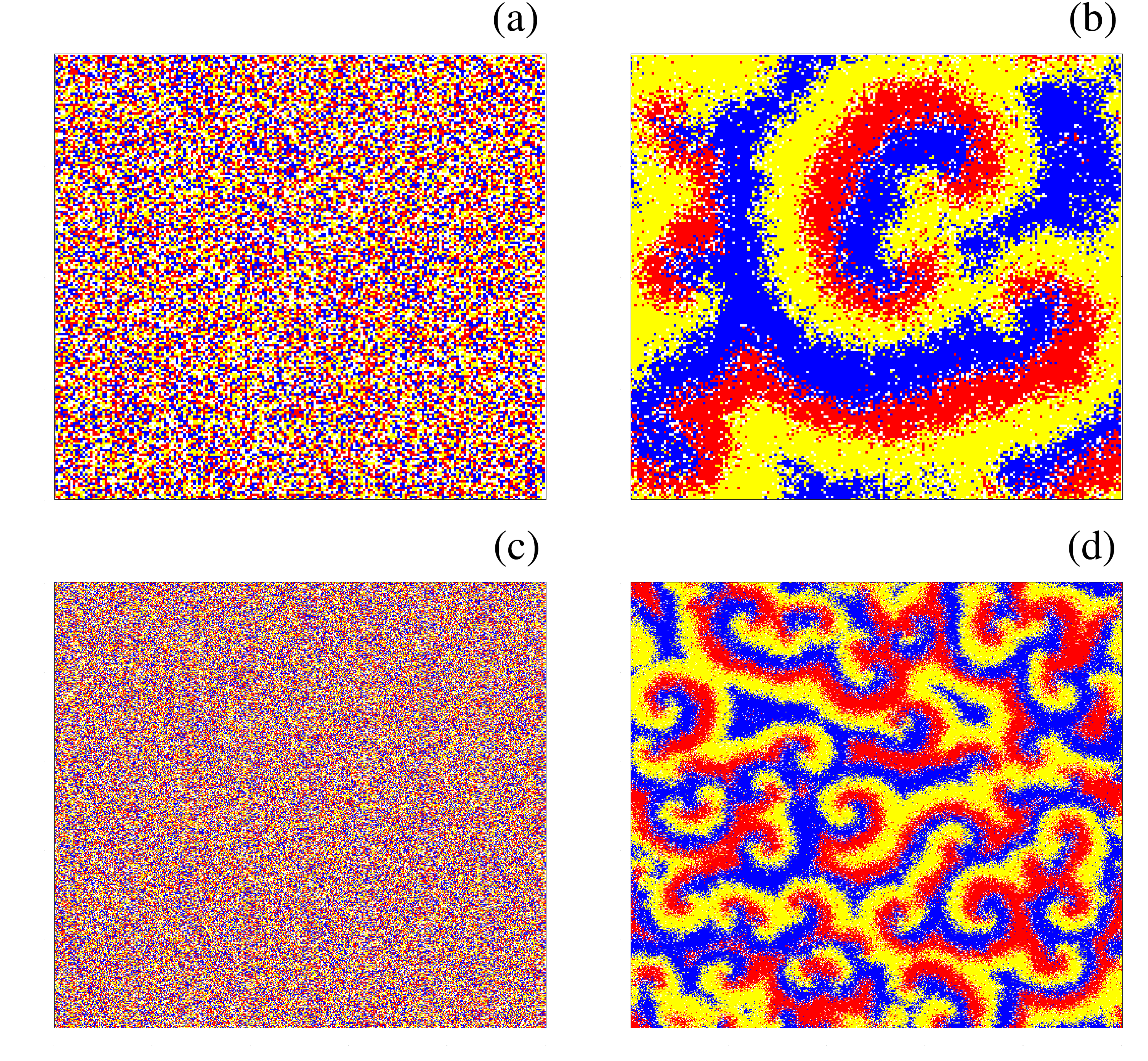}
\caption{The initial state for the lattice with (a) $N=200$ and (c) $N=500$ and a snapshot after $10000$ generations in (b) and (d), respectively. For both cases, $m=0.5$ and $p=r=0.25$.}
\label{fig1}
\end{figure}

We model the system's dynamics using stochastic simulations, which follow the standard formalism. The time evolution of the model is implemented via the use of random access to the species and the rules they obey. The procedure starts preparing the initial state, which is built as follows: we randomly select one site and one of the four possible states (species $a$, $b$, $c$ or an empty site $e$) which are identified by the colors red, blue, yellow and white, respectively, and paint the site with the corresponding color. This is repeated $N^2$ times to build the initial state. We then evolve the initial state by following the rules: we randomly select a site, which is the active site, and then a neighbor. We randomly select the rule and apply it: if the rule selected is mobility, the two sites exchange position, that is, under mobility $a\,b\to b\,a$, etc. If the rule is reproduction, we color the neighbor with the same color of the chosen site, if and only if the neighbor is empty, that is, under reproduction $a\,e\to a\,a$, $b\,e\to b\,b$, and $c\,e\to c\,c$, where $e$ represents an empty site. Finally, if the rule is competition, we use the rock-paper-scissors rules, that is $a\, b\to a\,e$, $b\,c\to b\,e$ and $c\,a\to c\,e$; see, e.g., Refs.~\cite{review1,review2,SR1,Ham,E1,E2,E4,E5,M1,M2,M3,ML}. We remark that the empty site is inert, that is, if the active site is empty, we return and choose another site. We do this $N^2$ times and this identifies the generation time. We account for the time evolution using the generation time as the appropriate unit of time. In Fig.~\ref{fig1}, we illustrate the time evolution of the stochastic simulations that we implement in this work with an initial state and a snapshot after 10000 generations, for lattices with $200\times200$ and $500\times500$ sites. The initial state is where all the four possible states are uniformly distributed, as we can see in Figs.~\ref{fig1} (a) and (c). At generation $g=10000$, we can observe spiral structures in the lattice, for $N=200$ and $N=500$, in Figs.~\ref{fig1} (b) and (d) respectively. The spiral structures are typical of systems that evolve under the above rules, with the three species competing cyclically and keeping biodiversity, with the abundance oscillating around an average, but never reaching zero or unit, to disappear or dominate the system, respectively. There are other interesting studies on the presence of stable and unstable spiral patterns in similar models; see. e.g., Refs.~\cite{AA,BB}.

\section{Hamming distance}\label{sec2}

Let us now revisit the Hamming distance density investigated before in \cite{SR1,Ham}. In the current context, the Hamming distance measures the difference between two $N \times N$ matrices. To implement this, we first consider the lattice that describes the initial state as an $N\times N$ matrix, and make a copy of it. The initial state is used to run the simulation to get to the final state, which is then saved. However, during the time evolution a new file is created, in which one saves every single step used to run it. One then takes the copy of the initial state and randomly selects a lattice site and modifies its content, changing its color to another one. This new state
has the tiniest difference, since among the $N^2$ lattice sites it has a single site which is different from the initial
state already used to evolve in time. We use this new initial state to run the same simulation already considered, evolving it according to the very same rules that appear in the saved file. The procedure leads to another final state, which is also saved. We then count the number of distinct sites between the two lattices in every generation, divide it by $N^2$ to get the Hamming distance density. This was studied before in \cite{SR1}, and more specifically in \cite{Ham}, where the measure was shown to have a universal qualitative behavior, despite its quantitative dependence on the initial state.

We also take the opportunity and investigate how the increase of mobility contributes to break biodiversity. This is an interesting issue, firstly suggested in \cite{C}, that was also examined in Refs.~\cite{review2,Ham,E1,E2,E4,E5,M1,M2,M3}, and we use this as a mechanism to help us to investigate the basin entropy behavior in the novel environment that we explore below. To illustrate the results, we investigate the Hamming distance density with $m = 0.5$, which is below the critical mobility, in the region that mantains the biodiversity. We also study the extinction case, with m = 0.98, which is above the critical value, so in the region that breaks diversity. We do not investigate the critical value of mobility, since it was already investigated in Refs.~\cite{C,review2,Ham,E1,E2,E4,E5,M1,M2,M3} and is now a known fact. 

\begin{figure}[t!]
\centering
\includegraphics[width=0.98\linewidth]{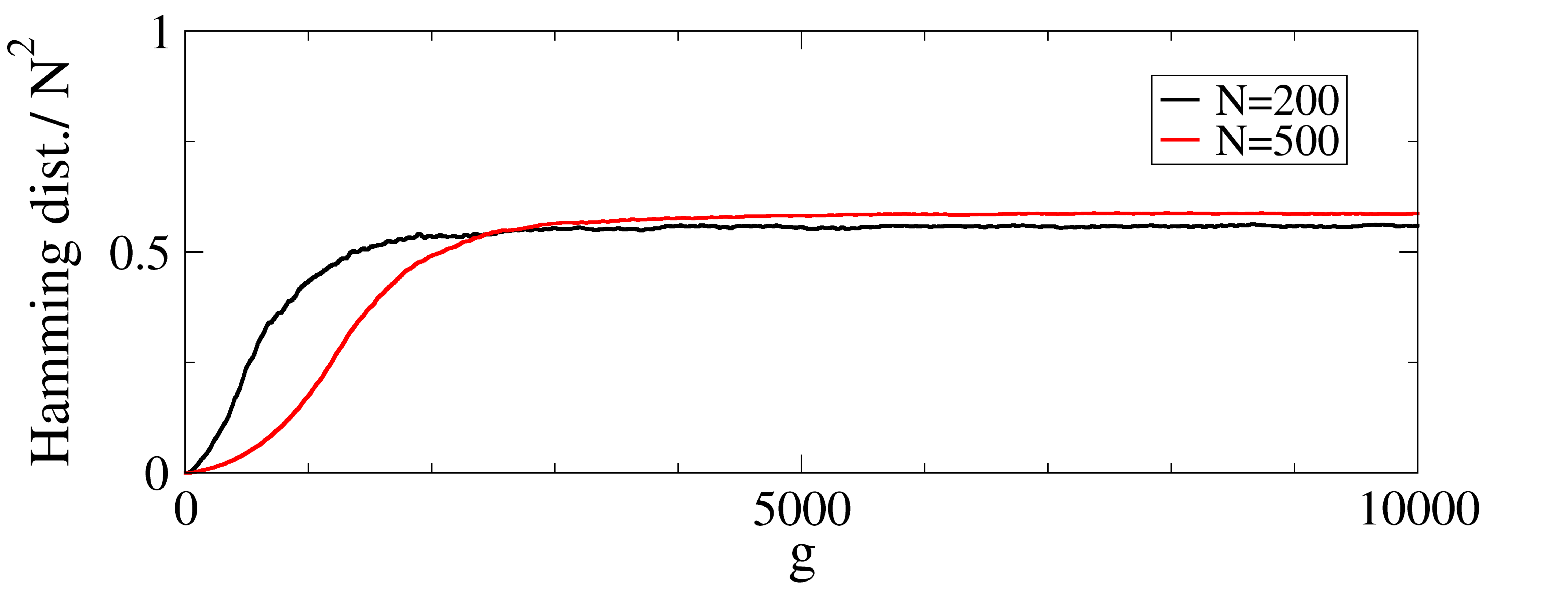}
\caption{The Hamming distance density, obtained in the lattice with size $200\times 200$ (black curve) and $500\times 500$ (red curve), for $10000$ generations. For both cases, $m=0.5$ and $p=r=0.25$.}
\label{fig2}
\end{figure}

The motivation to study the Hamming distance density is to compare its behavior with the novel investigation to be done in the next section, where we adapt the basin entropy concept to the network simulations that we develop in this work and use it to unveil the chaotic behavior it engenders. We describe the Hamming distance density running the simulations for two lattice sizes, with $N=200$ and $N=500$, and we evolve the simulations up to 10000 generations. The results of these simulations are displayed in Fig.~\ref{fig2} in the case the mobility is below the critical mobility, and from it we can observe the universal qualitative behavior unveiled before in \cite{SR1,Ham}: the Hamming distance density starts at the very small value $1/N^2$ and increases smoothly, stabilizing at some average value below unity. The results of Fig. \ref{fig2} are displayed as an average over $100$ realisations, starting from different initial states. The simulations are similar to the case investigated before in \cite{Ham}, so we omit further details here.

\begin{figure}[t!]
\centering
\includegraphics[width=0.98\linewidth]{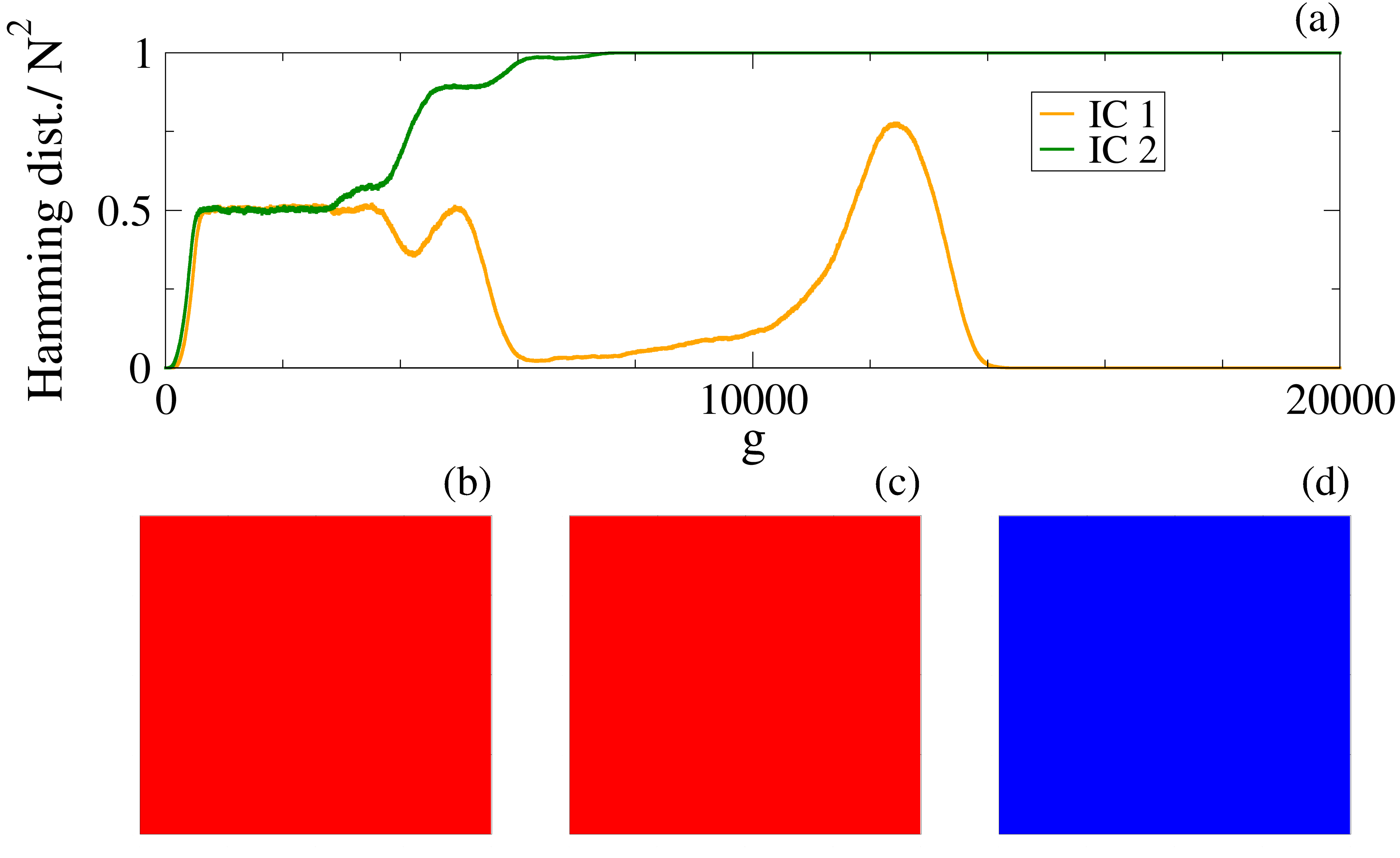}
\caption{The Hamming distance density for $m=0.98$ and different initial conditions in a lattice with size $200\times 200$. (a) The Hamming distance density is calculated between the evolution of the original grid and the evolution up to 20000 generations of two others initial condition: IC1 (orange curve)  and IC2 (green curve). In figure (b), (c) and (d) we have the final snapshot from the original initial condition, IC1 and IC2, respectively.}
\label{fig3}
\end{figure}

We also display in Fig.~\ref{fig3} (a) the Hamming distance density for the lattice with $N=200$, with the mobility very close to unit, so above the critical mobility. Here we run the network simulations up to $20000$ generations. In this case, the Hamming distance density is calculated between the evolution of the original grid and the evolution of two others initial condition: IC1  and IC2. The initial conditions IC1 and IC2 are equal to the original one, except from one site, randomly chosen. In Figs.~\ref{fig3} (b) and (c), the winner species is the red one, so, the Hamming distance becomes zero in Fig.~\ref{fig3} (a). From  Fig.~\ref{fig3} (d), the winner species is the blue one, so the Hamming distance becomes one, relative to the original one, represented in Fig.~\ref{fig3} (b). In both cases we note that diversity is broken for very high values of the mobility: if we change only one site the winner species can be one of the three different species, giving to the Hamming distance the value zero or one. These results are in good agreement with the previous works described in Ref.~\cite{Ham}.


\section{Basin entropy}\label{BE}

The basin entropy is a useful tool that can be applied to nonlinear dynamics to enable us to measure the final state unpredictability for numerical and experimental setups. Here we want to use this recently introduced concept of basin entropy \cite{SR2} to characterize the time evolution which we are studying in this work.

Since we are dealing with the stochastic network evolution of a cyclic three-species system, the basin entropy can be calculated in the following way. Firstly, we consider the lattice with size $N\times N$ and then we divide the lattice in $N_b\times N_b$ square non-overlapping boxes. For each box inside the lattice, we compute the Gibbs entropy given by

\begin{equation}\label{Si}
S_i=-\sum_{j=1}^{N_A}p_{ij}\log{p_{ij}}.
\end{equation}
In this work, we use $N_A=4$ because the set of possibilities in the lattice includes the three species and the empty site. The term $p_{ij}$ represents the proportion of how many sites inside the $i^{th}$ box are occupied by the specific species $j^{th}$. In the limit case, where all states inside the box are occupied by the same $j^{th}$ state, the contribution to the sum of Eq.~\ref{Si} is equal to zero. On the other hand, the entropy is maximum if all possible states are equiprobable, in this case its value is equal to $\log{N_A}$. 

The second step is to make the sum of all terms of entropy that cover the full lattice, that is, 
\begin{equation}
S=\sum_{i=1}^{L_b}S_i,
\end{equation}
where $L_b$ is the number of boxes of size $N_b\times N_b$ in the lattice $N\times N$. Finally, the basin entropy is defined by the expression 
\begin{equation} 
S_b={S}/{L_b}.
\end{equation}

\begin{figure}[t!]
\centering
\includegraphics[width=0.98\linewidth]{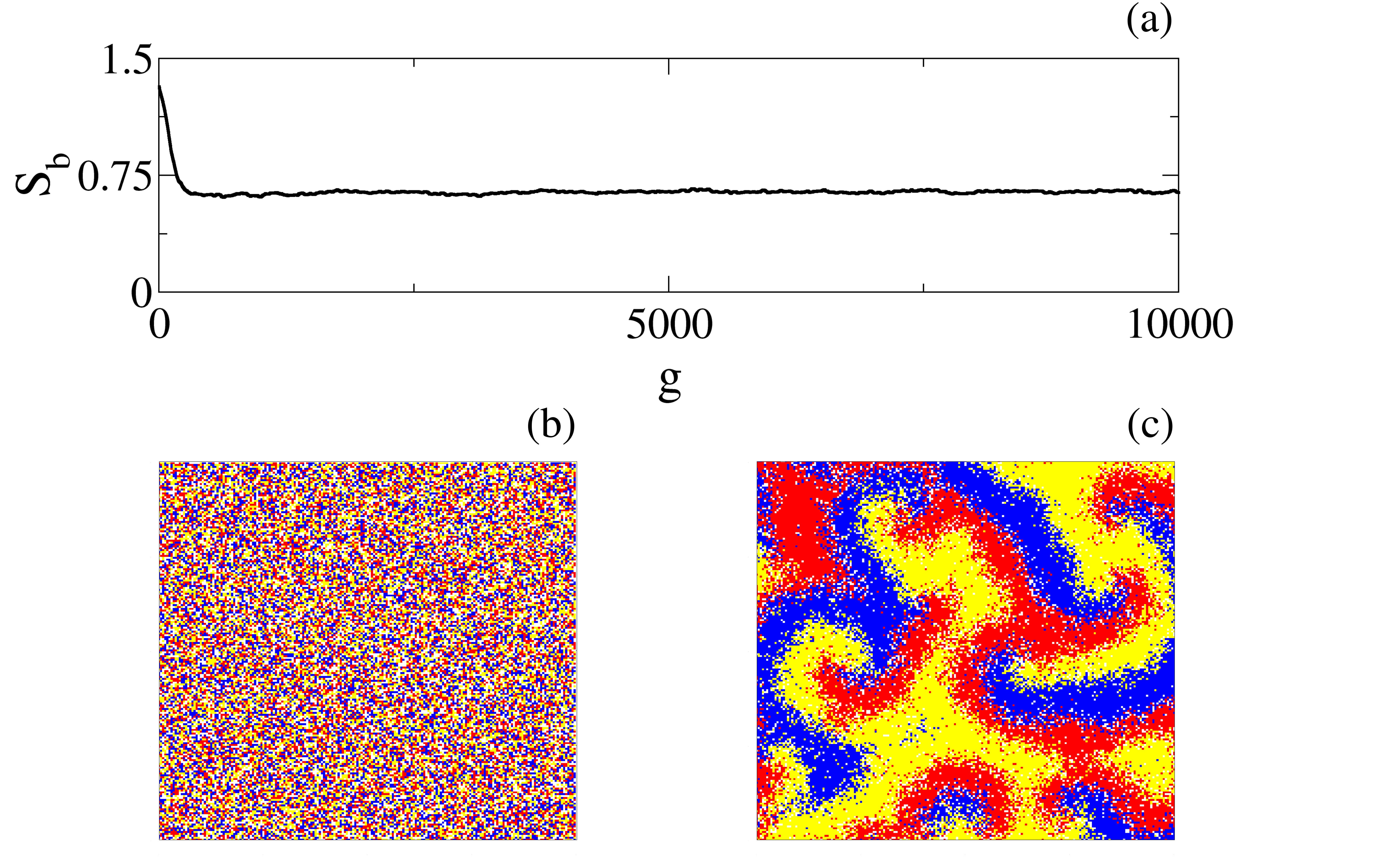}
\caption{(a) The basin entropy for the lattice with size $200\times 200$, $m=0.5$ and $p=r=0.25$. In the cases (b) and (c) we represent the lattices for the initial state and a snapshot for generation $g=10000$, respectively.}
\label{fig4}
\end{figure}

\begin{figure}[t!]
\centering
\includegraphics[width=0.98\linewidth]{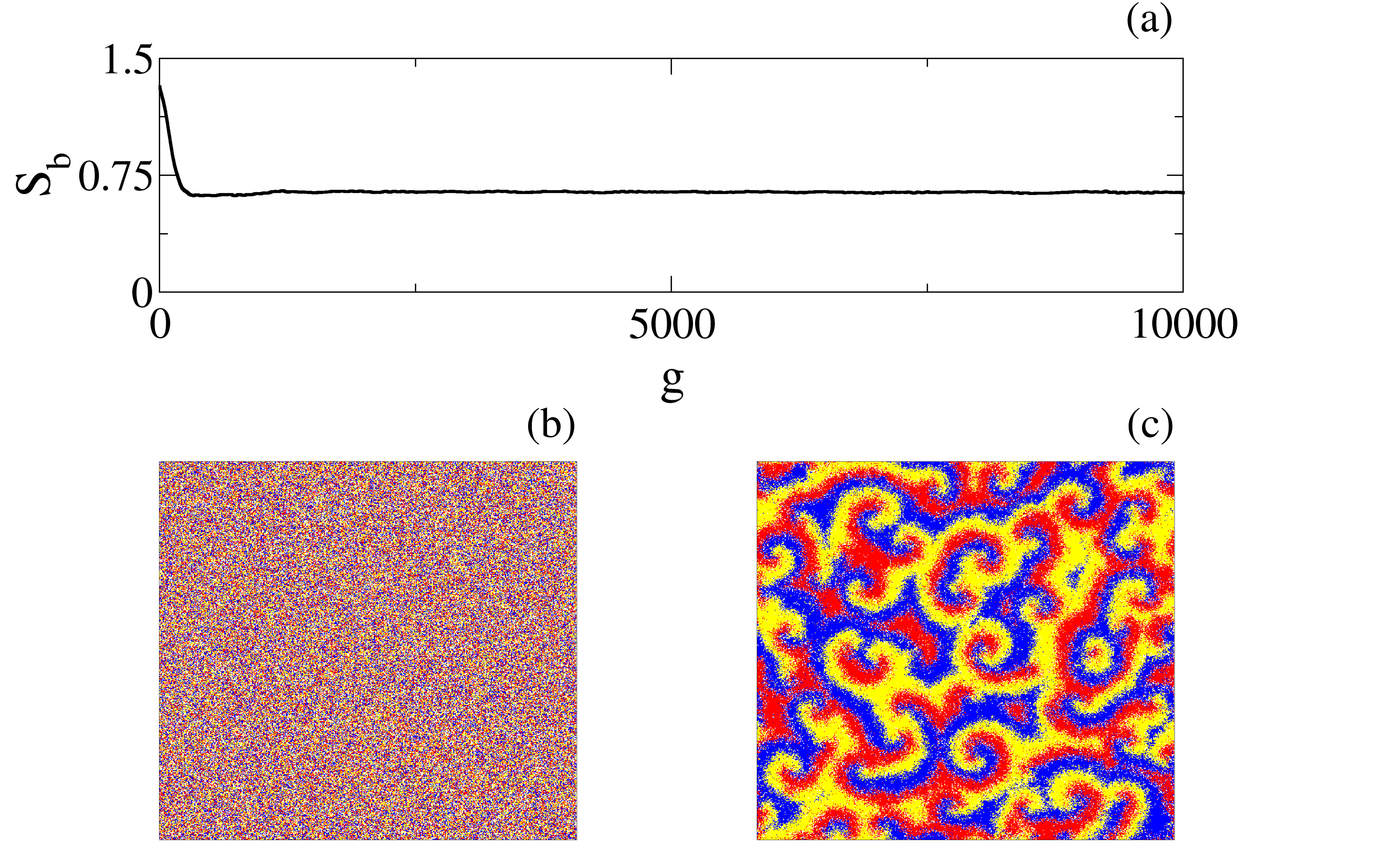}
\caption{(a) The basin entropy for the lattice with size $500\times 500$, $m=0.5$ and $p=r=0.25$.  The initial state is shown in (b). The grid for the generation $g=10000$ is represented in (c).}
\label{fig5}
\end{figure}

\begin{figure}[t!]
\centering
\includegraphics[width=0.98\linewidth]{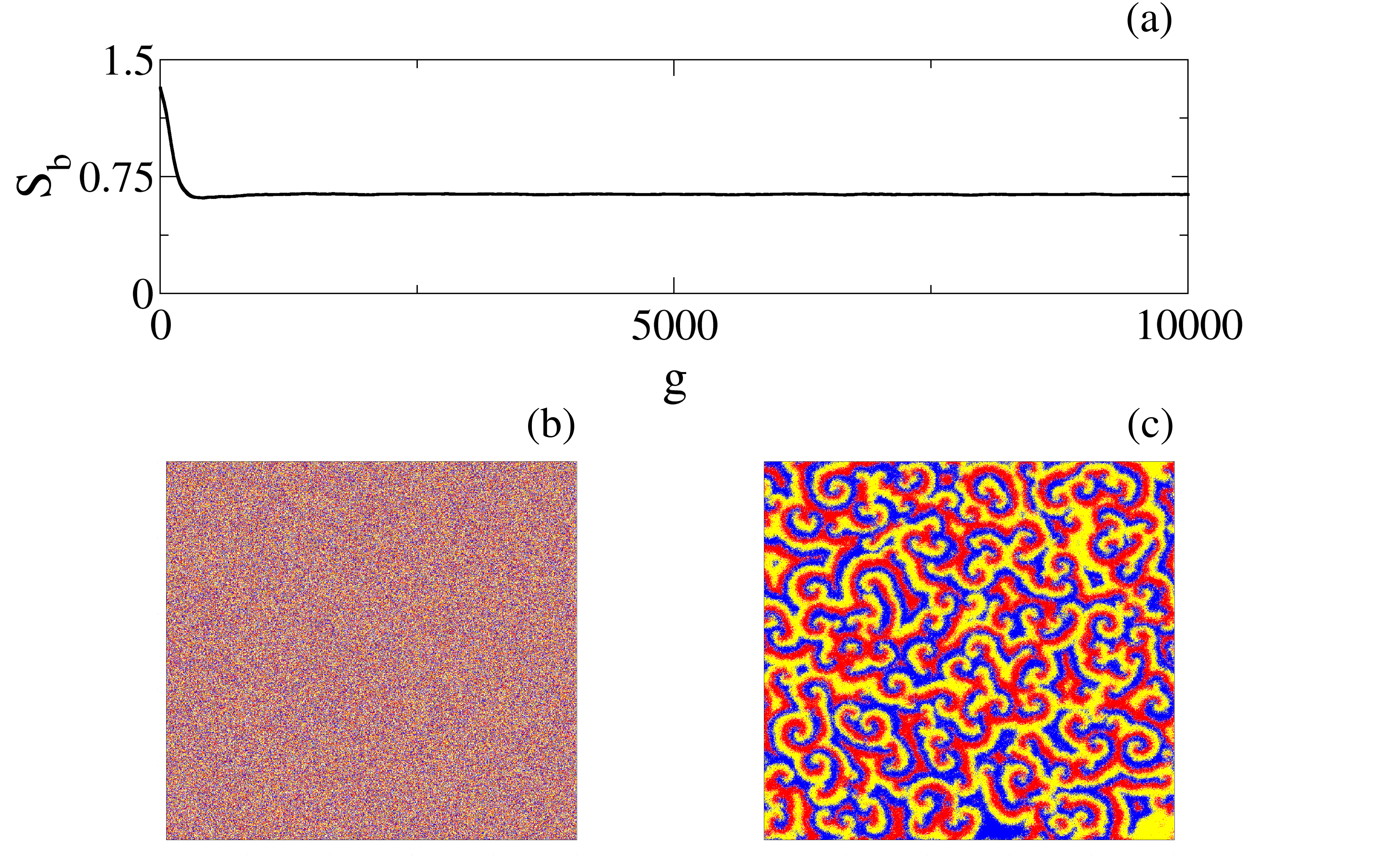}
\caption{(a) The basin entropy for the lattice with size $1000\times 1000$, $m=0.5$ and $p=r=0.25$.  The initial state is shown in (b). The snapshot for the generation $g=10000$ is depicted in (c).}
\label{fig6}
\end{figure}

Let us now implement the basin entropy simulations with a box of size $5 \times 5$. Moreover, to see how the basin entropy behaves as we increase the lattice size, we take three lattice sizes, with $N=200$, $500$, and $1000$. In Figs.~\ref{fig4}, \ref{fig5} and \ref{fig6} we depict the basin entropy for these three distinct lattices, running the simulations up to 10000 generations. In all cases, we used for the mobility the value $m=0.5$ (with $p=r=0.25$), which is below the critical mobility, so the systems keep biodiversity. Since the basin entropy depends on the initial state, the results shown in these figures were obtained from an average over 10 distinct simulations, each one started with a different initial state. We see from the results that the basin entropy $S_b$ starts at the highest possible value: an state where all the four possible states are equiprobable, as we can see in Figs.~\ref{fig4} (b), \ref{fig5} (b) and \ref{fig6} (b). We run the simulations and the systems evolve in time with the basin entropy decreasing and converging toward a positive value. The convergence is reached after the spiral structures are formed in the lattice, as we observe from the snapshots displayed in ~\ref{fig4} (c), \ref{fig5} (c) and \ref{fig6} (c). The positive values of $S_b$ that appear in Figs.~\ref{fig4} (a), \ref{fig5} (a), and \ref{fig6} (a) show that the basin entropy seems to be insensitive to the size of the lattice and suggest that the stochastic simulations that we are monitoring evolve unveiling a complex chaotic behavior. 

In order to further understand how the results shown in Figs.~\ref{fig4},
\ref{fig5} and \ref{fig6} depend on the size of the lattice, let us calculate $S_b$ using $m=0.5$ and $r=p=0.25$ for each one of the three lattice sizes, using an average over $20$ simulations, each one starting from a different initial state for the three distinct lattice sizes. Since the results in Figs.~\ref{fig4}, \ref{fig5} and \ref{fig6} show that $S_b$ converges rapidly, we calculate each $S_b$ at the time $g=3000$. The results are shown in Table 1, for the lattices with $200\times200$, $500\times500$, and $1000\times1000$ sites, respectively. They indicate that under the presence of biodiversity, the basin entropy is positive and independent of the lattice size.

\begin{table}[!h]\label{Tab}
\begin{center}
\begin{tabular}{|c|c|}
	\hline
	\footnotesize{\textbf{Lattice Size} } & \footnotesize{\textbf{Basin Entropy}} \\
	\hline
	\footnotesize{$200\times 200$} & \footnotesize{$0.63\pm0.02$}\\ 
	\hline
	\footnotesize{$500\times 500$} & \footnotesize{$0.64\pm0.01$} \\
	\hline
	\footnotesize{$1000\times 1000$} & \footnotesize{$0.637\pm0.003$}\\ 
	\hline
\end{tabular}
\end{center}
\caption{Average and standard deviation of the basin entropy for different lattice sizes using $20$ simulations. The values of the parameters are $m=0.5$ and $p=r=0.25$, with the time evolution ending at $g=3000$.}
\end{table}

To investigate the dependence of the basin entropy on the size of the box used in the simulations, let us now consider two other box sizes, one with $4\times 4$ and the other with $10\times 10$ sites. Although there is no qualitative difference, we noted that the basin entropy depends quantitatively on the size of the box used to implement the simulations. The results are shown in the Table 2, obtained at the time $g=3000$ with an average over $20$ distinct simulations. They corroborate the previous investigation, that the basin entropy does not depend on the size of the lattice. However, it depends on the size of the box used to simulate the entropy, and it increases as the size of the box is increased.

\begin{table}[!h]\label{Tab}
\begin{center}
\begin{tabular}{|c|c|}
	\hline
	\footnotesize{\textbf{Box Size} } & \footnotesize{\textbf{Basin Entropy}} \\
	\hline\hline
	\footnotesize{$4\times 4$} & \footnotesize{$0.60\pm0.03$} \\ 
	\hline
	\footnotesize{$5\times 5$} & \footnotesize{$0.63\pm0.02$} \\
	\hline
	\footnotesize{$10\times 10$} & \footnotesize{$0.75\pm0.03$} \\ 
	\hline\hline
	\footnotesize{$4\times 4$} & \footnotesize{$0.605\pm0.009$} \\ 
	\hline
	\footnotesize{$5\times 5$} & \footnotesize{$0.64\pm0.01$} \\
	\hline
	\footnotesize{$10\times 10$} & \footnotesize{$0.76\pm0.01$} \\ 
	\hline\hline
	\footnotesize{$4\times 4$} & \footnotesize{$0.600\pm0.005$} \\ 
	\hline
	\footnotesize{$5\times 5$} & \footnotesize{$0.637\pm0.003$} \\
	\hline
	\footnotesize{$10\times 10$} & \footnotesize{$0.750\pm0.006$} \\ 
	\hline
\end{tabular}
\end{center}
\caption{The basin entropy for the three distinct box sizes $4\times4$, $5\times5$ and $10\times10$. They are shown in three blocks of three lines, which account for the three lattice sizes, with $N=200$, $500$ and $1000$, from top to bottom, respectively.}
\end{table} 

\begin{figure}[t!]
\centering
\includegraphics[width=0.98\linewidth]{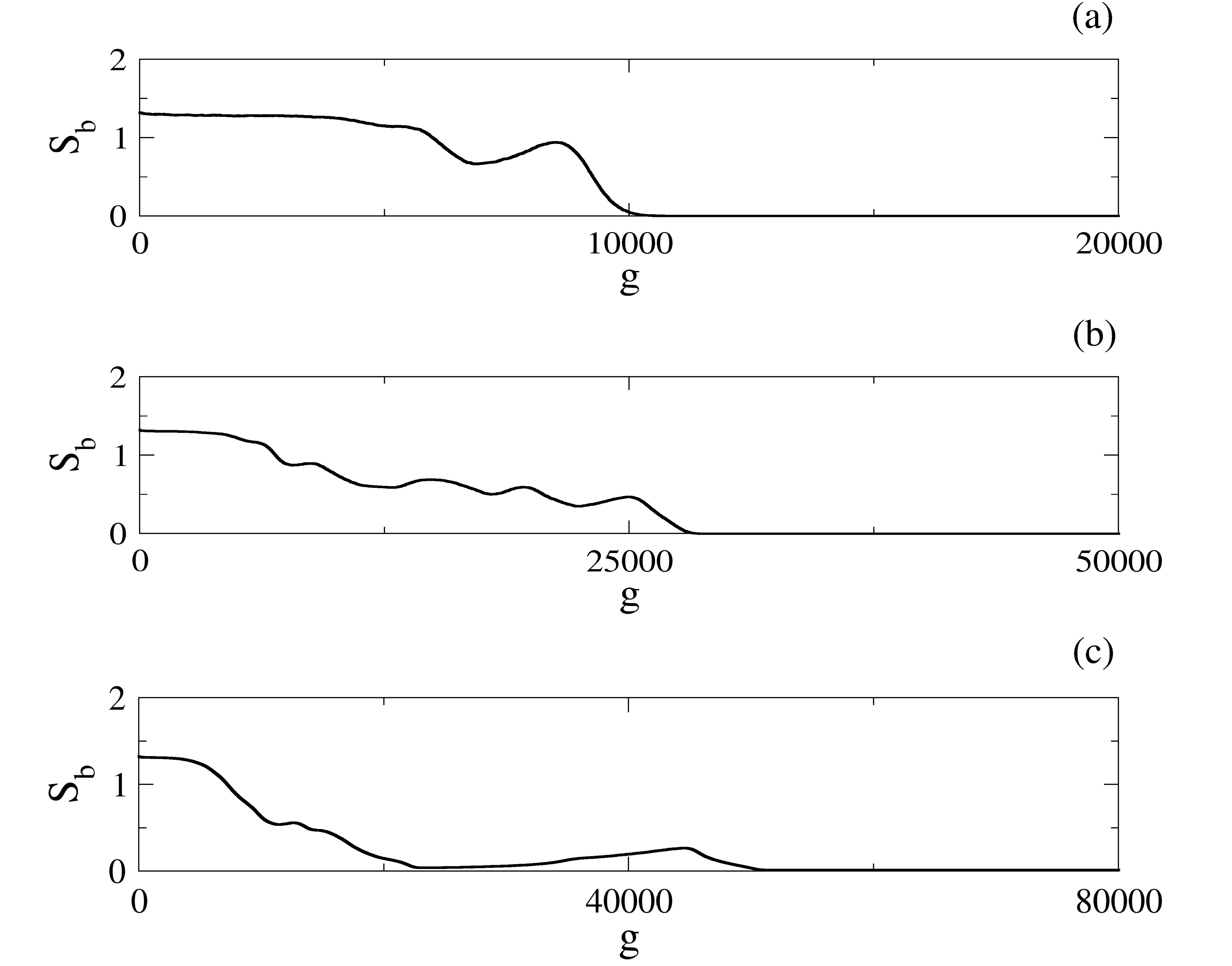}
\caption{The basin entropy for $m=0.99$ and $p=r=0.005$, in a lattice with size $200\times 200$ (a), $500\times500$ (b), and $1000\times1000$ (c). }
\label{fig7}
\end{figure}

In order to further stress the importance of the study of the basin entropy behavior, we take the same three lattices and the box of size $5\times5$, but now we calculate the basin entropy using for the mobility the value $m=0.99$ and $p=r=0.005$. This value is above the critical mobility, so the system is supposed to evolve toward a trivial final state containing a single species, ending biodiversity. We then expect that $S_b$ ends up vanishing as time goes by. The results are depicted in Fig.~\ref{fig7}, for the three distinct lattice sizes, for $g=20000$, $50000$ and $80000$, respectively. They show that the basin entropy goes to zero as the system evolves in time, indicating that in the long run the system always loses biodiversity. Although in Fig.~\ref{fig7} we are not showing the snapshots at that final state, we confirmed that the lattice is entirely filled with a single color which can be red, blue or yellow. We note, however, that although $S_b$ always vanishes, it takes longer times to vanish as the lattice size increases, showing that larger lattice sizes delay but do not prevent the extinction of biodiversity.

The above results show that, differently from the Hamming distance density, the basin entropy evolves in time decreasing from the value that measures the well-mixed initial state to another state in which the three species strive to aggregate in order to survive in the competitive environment. Moreover, the basin entropy has the advantage that it is easier to be implemented numerically, since one does not need to compare two distinct lattice evolutions at every generation, and also, it always goes to zero in the case of the end of biodiversity, independently of the initial state to be considered to implement the simulation.

We have done other simulations, for other values of mobility, but they all corroborate the results displayed in Figs.~\ref{fig4}, \ref{fig5}, and \ref{fig6}, when it is below the critical mobility, and in Fig.~\ref{fig7}, when it is above the critical mobility.


\section{Comments and conclusions}\label{sec4}

In this work we investigated the behavior of a model described by three distinct species that evolve in time governed by the action of mobility, competition and reproduction, with competition being controlled by the rules of the rock-paper-scissors game. In Fig.~\ref{fig1} we displayed the initial state and a snapshot after $10000$ generations, for the lattices with $200\times200$ and $500\times500$ sites, with the parameters given by $m=0.5$ and $r=p=0.25$. We then reviewed the Hamming distance behavior of the system for two distinct lattice sizes. We considered lattices with $200\times200$ and $500\times500$ sites, and the results that are shown in Figs.~\ref{fig2} and \ref{fig3} unveiled good agreement with the investigations reported before in Refs.~\cite{SR1,Ham}. 

We then considered a new possibility, that is, we used the basin entropy concept to describe the chaotic behavior of the system. In this new case, we considered three distinct lattices, with sizes $200\times200$, $500\times500$ and $1000\times1000$. The results showed that when mobility is not too high, the systems preserve biodiversity and the basin entropy decreases from a higher initial value to a lower but asymptotically constant and positive value. They are depicted in Figs.~\ref{fig4}, \ref{fig5} and \ref{fig6}. We also calculated the average value of the basin entropy to show that is is almost insensitive to the lattice size. In order to see how the basin entropy depends on the size of the box required to implement the simulation, we have also used three distinct box sizes, with $4\times4$, $5\times5$, and $10\times10$ sites. The results are shown in Table 1 and Table 2, and they indicate that for a given box size, the basin entropy does not depend on the lattice size. However, it increases as we increase the box size. Furthermore, we investigated the basin entropy for the very large value of mobility, $m=0.99$. In this case biodiversity is broken and, as a consequence, the basin entropy ends up vanishing, indicating that the system evolves to a trivial final state with a single species filling the entire lattice. These results are shown in Fig.~\ref{fig7}.

The results displayed in Figs.~\ref{fig4}, \ref{fig5}, \ref{fig6} and \ref{fig7} and in Table 1 and Table 2 allow to conclude that the basin entropy concept is another useful tool that can be used to investigate biodiversity in a three-species model that evolves cyclically under the rules of the rock-paper-scissors game. As we have shown, the basin entropy concept which we implemented in this work is simpler and faster to be implemented numerically, since it does not need to compare two distinct lattice evolutions which is required to get the Hamming distance. Although the simulations require the presence of boxes of size $N_b\times N_b$, this is not a problem if one chooses $N_b$ as a submultiple on $N$, bigger than the number of species but much smaller than $N$ itself.

Since the current investigation is the first study on the subject, further research is welcome and we are now exploring how the Hamming distance and basin entropy behave in generalized models, when one adds more species and/or modifies the rules that control the time evolution, changing the dynamics of the system; see, e.g., the models investigated in Refs.~\cite{AA,BB}. In particular, we are also interested in the study of the basin entropy under the presence of an apex predator, in a model similar to the case recently considered in Ref.~\cite{AP}. More specifically, the investigation implemented in the recent work \cite{bos} unveiled that the apex predator decaying parameter can be used to control the time evolution of the system, which may terminate into three qualitatively different possibilities, separated by two distinct phase transitions. It is of current interest to study how the Hamming distance and the basin entropy behave as one varies the apex predator decaying parameter, reaching and crossing the two critical phase transition values. We hope to report on this and in other related issues in the near future.

\acknowledgments{This study was financed in part by Conselho Nacional de Desenvolvimento Cient\'\i fico e Tecnol\'ogico (CNPq), Coordena\c{c}\~{a}o de Aperfei\c{c}oamento de Pessoal de N\'{i}vel Superior (CAPES) and Funda\c{c}\~{a}o Arauc\'{a}ria (Brazilian agencies). FMA acknowledges support from the CNPq grant 313274/2017-7, JDS acknowledges support from the CNPq grant 310124/2017-4 and DB acknowledges support from the CNPq grant 306614/2014-6. JDS and MM also acknowledge valuable discussions with Prof. RL Viana.}


\begin{thebibliography}{99}
\bibitem{A}B. Kerr, M.A. Riley, M.W. Feldman, and B.J.M. Bohannan, Nature {418}, 171 (2002).
\bibitem{B}B.C. Kirkup and M.A. Riley, Nature {\bf428}, 412 (2004).
\bibitem{C}T. Reichenbach, M. Mobilia, and E. Frey, Nature {\bf448}, 1046 (2007).
\bibitem{review1}G. Szab\'o and G. F\'ath, Physics Reports {\bf446}, 97 (2007).
\bibitem{review2}A. Szolnoki, M. Mobilia, L.-L. Jiang, B. Szczesny, A.M. Rucklidge, and M. Perc,  J. Royal Society Interface 11, 20140735 (2014).
\bibitem{SR1}D. Bazeia, M.B.P.N. Pereira, A.V. Brito, B.F. de Oliveira, and J.G.G.S. Ramos, Scientific Reports {\bf7}, 44900 (2017).
\bibitem{Ham}D. Bazeia, J. Menezes, B.F. de Oliveira, and J.G.G.S. Ramos, EPL {\bf119}, 58003 (2017).
\bibitem{SR2}A. Daza, A. Wagemakers, B. Georgeot, D. Gu\'ery-Odelin, and M.A.F. Sanju\'an, Scientific Reports {\bf6}, 31416 (2016).
\bibitem{PRE}M. Mugnaine, A.C. Mathias, M.S. Santos, A.M. Batista, J. D. Szezech, Jr., and R.L. Viana, Phys. Rev. E {\bf97}, 012214 (2018). 
\bibitem{PRA}A. Daza, B. Georgeot, D. Gu\'ery-Odelin, A. Wagemakers, and M.A.F. Sanju\'an, Phys. Rev. A {\bf95}, 013629 (2017).
\bibitem{M1}S. Venkat and M. Pleimling, Phys. Rev. E {\bf81} 021917 (2010).
\bibitem{E1}J. Juul, K. Sneppen and J. Mathiesen, Phys. Rev. E {\bf87}, 042702 (2013).
\bibitem{E2}J. Vukov, A. Szolnoki, and G. Szab\'o, Phys. Rev. E {\bf88}, 022123 (2013).
\bibitem{M2}H. Cheng et al., Scientific Report {\bf4}, 7486 (2014).
\bibitem{E4}A. Szolnoki and M. Perc, Phys. Rev. E {\bf93}, 062307 (2016).
\bibitem{M3}M. Mobilia, A.M. Rucklidge, and B. Szczesny, Games {\bf7}, 24 (2016).
\bibitem{E5}A. Szolnoki and M. Perc, Scientific Reports {\bf6}, 38608 (2016).
\bibitem{ML}R. May and W. Leonard, SIAM J. Appl. Math. {\bf29}, 243 (1975).
\bibitem{AA}B. Szczesny, M. Mobilia, and A. Rucklidge, EPL {\bf102}, 28012 (2013).
\bibitem{BB}B. Szczesny, M. Mobilia, and A. Rucklidge, Phys. Rev. E {\bf90}, 032704 (2014).
\bibitem{AP}C.A. Souza-Filho, D. Bazeia, and J.G.G.S. Ramos, Phys. Rev. E {\bf95}, 062411 (2017).
\bibitem{bos}D. Bazeia, B.F. de Oliveira, and A. Szolnoki, EPL {\bf124}, 68001 (2018).
\end{thebibliography}
\end{document}